\documentclass[jkps,twoside,twocolumn,fleqn,showpacs,showkeys,showdoi]{revtex4}
\usepackage{graphicx}
\usepackage{color}
\usepackage{amssymb}
\usepackage{amsmath}
\usepackage{bm}
\usepackage{eurosym}
\usepackage{enumerate}

\newcommand{\be}[1]{\begin{equation}\label{eq:#1}}
\newcommand{\ee}{\end{equation}}
\newcommand{\bea}{\begin{eqnarray}}
\newcommand{\eea}{\end{eqnarray}}
\newcommand{\bt}{\textbf}

\newcommand{\phd}{\phantom{\dag}}
\newcommand{\ph}{\phantom{.}}

\newcommand{\up}{^{\phd}}
\newcommand{\noi}{\noindent}
\newcommand{\no}{\nonumber}

\newenvironment{changemargin}[2]{\begin{list}{}{\setlength{\topsep}{0pt}  \setlength{\leftmargin}{#1}  \setlength{\rightmargin}{#2}
\setlength{\listparindent}{\parindent}
\setlength{\itemindent}{\parindent}  \setlength{\parsep}{\parskip}
}\item[]}{\end{list}}

\newcommand{\lcm}{\begin{changemargin}{-1.3cm}{0.5cm}}
\newcommand{\ecm}{\end{changemargin}}
\newcommand{\tbcm}{\begin{changemargin}{-0.8cm}{-0.25cm}}
\newcommand{\tecm}{\end{changemargin}}

\begin{document}
\setcounter{page}{0}
\title[]{Engineering and manipulating topological qubits in 1D quantum wires}
\author{Panagiotis \surname{Kotetes}}
\email{panagiotis.kotetes@kit.edu}
\author{Gerd \surname{Sch\"{o}n}}
\affiliation{Institut f\"{u}r Theoretische Festk\"{o}rperphysik, Karlsruhe Institute of Technology, 76128 Karlsruhe, Germany}
\author{Alexander \surname{Shnirman}}
\affiliation{Institut f\"{u}r Theorie der Kondensierten Materie, Karlsruhe Institute of Technology, 76128 Karlsruhe, Germany}

\begin{abstract}
We investigate the Josephson effect in TNT and NTN junctions, consisting of topological (T) and normal (N) phases of semiconductor-superconductor 1D heterostructures in the presence of a 
Zeeman field. A key feature of our setup is that, in addition to the variation of the phase of the superconducting order parameter, we allow the orientation of the magnetic field to 
change along the junction. We find a novel magnetic contribution to the Majorana Josephson coupling that permits the Josephson current to be tuned by changing the orientation of the 
magnetic field along the junction. We also predict that a spin current can be generated by a finite superconducting phase difference, rendering these materials potential candidates for 
spintronic applications. Finally, this new type of coupling not only constitutes a unique fingerprint for the existence of Majorana bound states but also provides an alternative 
pathway for manipulating and braiding topological qubits in networks of wires.
\end{abstract}

\pacs{73.63.Nm, 74.78.-w, 03.67.Lx}

\keywords{Quantum wires, Majorana bound states, Josephson effect, Quantum computing}

\maketitle

\section{INTRODUCTION}

The recent indications in experiments with semiconductor-superconductor 1D he\-te\-ro\-structures \cite{Mourik,Deng,Furdyna} of the existence of Majorana bound states (MBS)
\cite{Majorana,Volovik zero modes,Read,Kitaev unpaired,Yakovenko,Ivanov,Sau proximity,Sau,Sau Top. phase transition,Oreg,Wimmer,Neupert,Choy magnetic nano,Rosch,Tsvelik,Tsvelik PDW,
Flensberg spiral,Simon,Wang SC+AFM,Martin} have also intensified the pursuit of designing a quantum computer based on topologically protected qubits \cite{Kitaev fault tolerant,Quantum 
Computing review}. The existence of a pair of spatially separated MBS in these systems will signal the appearance of a topological qubit. Here, the two-level system arises from the 
emergence of a zero-energy quasiparticle leading to a doubly degenerate ground state. The topological character of the qubit reflects an intrinsic particle-hole symmetry of the 
specific system, which provides the protection against external sources of decoherence and noise as long as neither a gap closes nor quasiparticles are excited \cite{Sau robustness,
Sau conserving,Bolech1,Chamon decay,Lim,Budich}. 

The potential applications of this class of topological qubits require, on one hand, finding unique fingerprints that unambiguously confirm their existence and, on the other hand, 
developing techniques that permit their manipulation with an eye to quantum information processing \cite{Alicea,Stone,Teo braidless,Sau universal TQC,Clarke controlling,
Flensberg charge TQC,Sau Majo exchange,Flensberg spin qubit,Bonderson,Halperin,Romito,Xue}. The majority of previous proposals for their detection \cite{Bolech2,Tewari2,Kraus,
Akhmerov electrically det Majo,Flensberg tunnneling,Sau interferometry,Fulga,Akmerov conductance,Maciejko noise} rely on tunne\-ling and transport features of the zero-energy fermionic 
excitation. More recently, alternative experimental routes have been explored, based on unconventional Josephson signatures \cite{Jiang}, distinct from the well-known $4\pi$-periodic 
Josephson effect typical for MBS \cite{Kitaev unpaired,Prada,Platero,Fazio}. 

\begin{figure}[t]
\includegraphics[width=5.5cm]{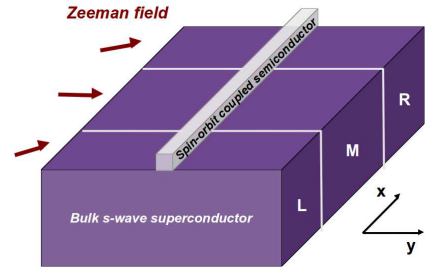}
\caption{(Color online) Platform for realizing topological qubits. A 1D semiconducting quantum wire with strong spin-orbit coupling is placed on top of a bulk s-wave superconductor. 
Proximity effects induce superconducting pairing to the quantum wire. Majorana bound states appear when we additionally apply a Zeeman field. The junctions that we consider should be 
fabricated by this type of hybrid structures.}
\label{fig1}
\end{figure}

In this manuscript, we report on new results concerning the Josephson effect in double-junctions of the type ${\rm TNT}$ and ${\rm NTN}$, consisting of  topo\-lo\-gi\-cal (${\rm T}$) 
and normal (${\rm N}$) phases. Each of the segments of the junction are constructed from heterostructures of 1D quantum wires (e.g. InSb) deposited on top of a bulk s-wave 
superconductor (e.g. Nb), in the simultaneous presence of a Zeeman magnetic field (Figure \ref{fig1}). We demonstrate that there exists a contribution to the Josephson coupling of 
these systems which can be tuned by varying the orientation of the Zeeman magnetic field along the junction. For two neighbouring MBS, this novel ``magnetic'' Josephson coupling in 
its simplest version is proportional to 
\bea
{\cal H}_{M}^{mag}\propto i\gamma_l\gamma_r\cos\left(\frac{\vartheta_l-\vartheta_r}{2}\right)\cos\left(\frac{\varphi_l-\varphi_r}{2}\right)\,,
\eea

\noi where $\gamma_{l,r}$ correspond to the Majorana operators satisfying $\{\gamma_{l,r},\gamma_{l,r}\}=1$ and $\{\gamma_{l},\gamma_{r}\}=0$, $\vartheta_{l,r}$ correspond to the 
angles defining the orientation of the Zeeman field and $\varphi_{l,r}$ represent the superconducting order parameter phases that the MBS see. 

We observe a $4\pi$-periodicity also for the Zeeman field phase difference $\vartheta_l-\vartheta_r$, that could provide a characteristic signature of the interacting MBS. The recent 
experimental results \cite{Furdyna} supporting the observation of the MBS Josephson effect in these he\-te\-rostructures, suggest that our setup constitutes a realistic, unambiguous 
and accessible way to detect the possible existence of MBS. Moreover, this extra term may offer alternative routes for braiding and manipulating MBS in networks of quantum wires, as 
for example in Y-junctions \cite{Sau Majo exchange,Halperin}, by adiaba\-ti\-cally varying the relative phases of the Zeeman field.


\section{Model Hamiltonian and bulk single particle eigenfunctions}

In order to study the Josephson effect of the junctions, we model the semiconductor-superconductor 1D he\-te\-ro\-structure with the following Hamiltonian
\bea
{\cal H}=\frac{1}{2}\int dx\ph\widehat{\Psi}^{\dag}(x)\widehat{{\cal H}}(\hat{p},x)\widehat{\Psi}(x)\,,\label{eq:H}
\eea

\noi where we have introduced the Bogoliubov - de Gennes Hamiltonian density
\bea
\widehat{{\cal H}}(\hat{p},x)&=&v\hat{p}\sigma_z+|\bm{{\cal B}}(x)|e^{-i\vartheta(x)\tau_z\sigma_z}\tau_z\sigma_x\no\\
&-&|\bm{\Delta}(x)|e^{-i\varphi(x)\tau_z}\tau_y\sigma_y+\left(\frac{\hat{p}^2}{2m}-\mu\right)\tau_z\,,\qquad\label{eq:Hdensity}
\eea

\noi and the enlarged Nambu spinor \bea\widehat{\Psi}^{\dag}(x)=\left(\begin{array}{cccc}\psi_{\uparrow}^{\dag}(x)&\psi_{\downarrow}^{\dag}(x)&\psi_{\uparrow}\up(x)&
\psi_{\downarrow}\up(x)\end{array}\right)\,.\eea 

\noi The factor $1/2$ in front of the Hamiltonian in Eq.(\ref{eq:H}) accounts for the doubling of the degrees of freedom when we introduce the extended Nambu spinor, which is composed 
by electronic creation and annihilation operators $\psi_{\sigma}^{\dag}(x)$ and $\psi_{\sigma}\up(x)$ of spin projection $\sigma=\uparrow,\downarrow$. To represent the possible terms 
of the Hamiltonian within this Nambu formalism, we have made use of Kronecker products of the Nambu-space $\bm{\tau}$ and spin-space $\bm{\sigma}$ Pauli matrices. The velocity $v$ 
corresponds to the strength of the spin-orbit interaction that is oriented along the $z-$axis, $\bm{{\cal B}}(x)$ represents the Zeeman magnetic field that is allowed to rotate freely 
in the $x-y$ plane, $\bm{\Delta}(x)=\left(\Delta_{\Re}(x),\Delta_{\Im}(x)\right)$ defines the proximity induced superconducting order parameter of the quantum wire written in a 
convenient vectorial form and finally the last term corresponds to the kinetic energy of the electrons measured from the chemical potential $\mu$.

For a bulk system, $\bm{{\cal B}}(x)$ and $\bm{\Delta}(x)$ become homogeneous and we can Fourier transform Eq.(\ref{eq:Hdensity}), obtaining
\bea
\widehat{{\cal H}}(k)&=&v\hbar k\sigma_z+|\bm{{\cal B}}|e^{-i\vartheta\tau_z\sigma_z}\tau_z\sigma_x-|\bm{\Delta}|e^{-i\varphi\tau_z}\tau_y\sigma_y\no\\
&+&\left(\frac{\hbar^2k^2}{2m}-\mu\right)\tau_z\,.\qquad\label{eq:HdensityFourier}
\eea

\noi Correspondingly, the extended Nambu spinor becomes $\widehat{\Psi}_k^{\dag}=(\begin{array}{cccc}\psi_{k\uparrow}^{\dag}&\psi_{k\downarrow}^{\dag}&\psi_{-k\uparrow}\up&
\psi_{-k\downarrow}\up\end{array})$. The Bo\-go\-liu\-bov operators that diagonalize the above Hamiltonian, are generally of the form 
$\gamma_k\up=u_{k\uparrow}\up\psi_{k\uparrow}\up+u_{k\downarrow}\up\psi_{k\downarrow}\up+v_{k\uparrow}\up\psi_{-k\uparrow}^{\dag}+v_{k\downarrow}\up\psi_{-k\downarrow}^{\dag}$. 
Majorana operators, satisfying $\gamma_k\up=\gamma_k^{\dag}$, may only occur for inversion symmetric momentum space points ($k\equiv-k$) for which we may have linear combinations of 
the type $\psi_{k\sigma}\up\pm\psi_{-k\sigma}^{\dag}\equiv\psi_{k\sigma}\up\pm\psi_{k\sigma}^{\dag}$ with $\sigma=\uparrow,\downarrow$. In the case of a bulk system this can take place 
only for $k=0$. In fact, the parameter regime where the energy spectrum $E(k)$ shows zeroes for $k=0$, indicates a phase boundary between the topological trivial and non-trivial phases. 
For the Hamiltonian of Eq.(\ref{eq:HdensityFourier}), one directly obtains that for $|\bm{{\cal B}}|>\sqrt{|\bm{\Delta}|^2+\mu^2}$ the system is in the topological phase with one MBS 
for $k=0$, while for $|\bm{{\cal B}}|<\sqrt{|\bm{\Delta}|^2+\mu^2}$ the system is in the topologically trivial phase with a zero or an even (in general) number of MBS 
\cite{Oreg,Alicea}.

In our case we set the chemical potential equal to zero $\mu=0$, which permits us to consider a truncated version of the Hamiltonian of Eq.(\ref{eq:HdensityFourier}). Specifically, we 
drop the kinetic energy term and our truncated Hamiltonian reads
\bea
\widehat{{\cal H}}_{tr}(k)=v\hbar k\sigma_z+|\bm{{\cal B}}|e^{-i\vartheta\tau_z\sigma_z}\tau_z\sigma_x-|\bm{\Delta}|e^{-i\varphi\tau_z}\tau_y\sigma_y\,.\phd
\label{eq:HdensityFourierTrunc}
\eea

\noi Considering the latter Hamiltonian in order to study the Josephson effect is naturally justified, since we want to examine topological properties that are related to the $k=0$ 
point. The truncated Hamiltonian is the linearized version of our initial one, about $k=0$. Nevertheless, as already discussed in Ref. \cite{Jiang} the models of 
(\ref{eq:HdensityFourier}) and (\ref{eq:HdensityFourierTrunc}) show a discrepancy concerning the characterization of the topological and normal phases. Specifically, in the linearized 
model the topological phase occurs when $|\bm{\Delta}|>|\bm{{\cal B}}|$ and the normal when $|\bm{\Delta}|<|\bm{{\cal B}}|$.

For studying the junctions we shall consider that the superconducting order parameter and the magnetic field remain constant for each segment, having the following spatial profile 
$\bm{{\cal B}}(x)=\bm{{\cal B}}_l+\left(\bm{{\cal B}}_m-\bm{{\cal B}}_l\right)\Theta(x-x_a)+\left(\bm{{\cal B}}_r-\bm{{\cal B}}_m\right)\Theta(x-x_b)$ and $\bm{\Delta}(x)=
\bm{\Delta}_l+\left(\bm{\Delta}_m-\bm{\Delta}_l\right)\Theta(x-x_a)+\left(\bm{\Delta}_r-\bm{\Delta}_m\right)\Theta(x-x_b)$, where $\Theta(x)$ is the Heaviside function and the label 
$s=l,m,r$ denotes the left, middle and right segments. Every segment can be described by the Hamiltonian of Eq.(\ref{eq:HdensityFourierTrunc}). Therefore it is eligible to determine 
the bound state eigenfunctions for the latter Hamiltonian. Since we are looking for bound states, we set $\kappa=ik$. The 4-component single particle bulk wavefunctions 
$\Psi(x)=e^{\kappa x}\Psi(\kappa)$ are readily obtained and read

\begin{widetext}
\bea
\left|\kappa_1(E);\bm{{\cal B}},\bm{\Delta}\right>&=&\frac{1}{\sqrt{2}}\left(\begin{array}{cccc}
i\sin\left(\frac{\omega_+}{2}\right)e^{-i\frac{\varphi+\vartheta}{2}},&-\cos\left(\frac{\omega_+}{2}\right)e^{-i\frac{\varphi-\vartheta}{2}},&
-i\sin\left(\frac{\omega_+}{2}\right)e^{+i\frac{\varphi+\vartheta}{2}},&\ph-\cos\left(\frac{\omega_+}{2}\right)e^{+i\frac{\varphi-\vartheta}{2}}\end{array}\right)^T\,,\qquad\\
\left|\kappa_2(E);\bm{{\cal B}},\bm{\Delta}\right>&=&\frac{1}{\sqrt{2}}\left(\begin{array}{cccc}
i\cos\left(\frac{\omega_+}{2}\right)e^{-i\frac{\varphi+\vartheta}{2}},&+\sin\left(\frac{\omega_+}{2}\right)e^{-i\frac{\varphi-\vartheta}{2}},&
-i\cos\left(\frac{\omega_+}{2}\right)e^{+i\frac{\varphi+\vartheta}{2}},&\ph+\sin\left(\frac{\omega_+}{2}\right)e^{+i\frac{\varphi-\vartheta}{2}}\end{array}\right)^T\,,\qquad\\
\left|\kappa_3(E);\bm{{\cal B}},\bm{\Delta}\right>&=&\frac{1}{\sqrt{2}}\left(\begin{array}{cccc}
\ph\sin\left(\frac{\omega_-}{2}\right)e^{-i\frac{\varphi+\vartheta}{2}},&-i\cos\left(\frac{\omega_-}{2}\right)e^{-i\frac{\varphi-\vartheta}{2}}s_-,&
\sin\left(\frac{\omega_-}{2}\right)e^{+i\frac{\varphi+\vartheta}{2}},&+i\cos\left(\frac{\omega_-}{2}\right)e^{+i\frac{\varphi-\vartheta}{2}}s_-\end{array}\right)^T\,,\qquad
\label{eq:wavefunction3}\\
\left|\kappa_4(E);\bm{{\cal B}},\bm{\Delta}\right>&=&\frac{1}{\sqrt{2}}\left(\begin{array}{cccc}
\ph\cos\left(\frac{\omega_-}{2}\right)e^{-i\frac{\varphi+\vartheta}{2}},&+i\sin\left(\frac{\omega_-}{2}\right)e^{-i\frac{\varphi-\vartheta}{2}}s_-,&
\cos\left(\frac{\omega_-}{2}\right)e^{+i\frac{\varphi+\vartheta}{2}},&-i\sin\left(\frac{\omega_-}{2}\right)e^{+i\frac{\varphi-\vartheta}{2}}s_-\end{array}\right)^T\,,\qquad
\label{eq:wavefunction4}
\eea 
\end{widetext}

\noi where we have made use of the definitions $\omega_{\pm}\equiv i\alpha_{\pm}+\pi/2$, $\tanh\alpha_{\pm}=E/||\bm{\Delta}|\pm|\bm{{\cal B}}||$, and 
$s_-\equiv{\rm sign}(|\bm{\Delta}|-|\bm{{\cal B}}|)$. The above wave-functions are characterized by the corresponding ``wave-vectors'' $\kappa_1(E)\equiv+\kappa_+(E)$, 
$\kappa_2(E)\equiv-\kappa_+(E)$, $\kappa_3(E)\equiv+\kappa_-(E)$ and $\kappa_4(E)\equiv-\kappa_-(E)$ where
\bea
\kappa_{\pm}(E)=\sqrt{\left(|\bm{\Delta}|\pm|\bm{{\cal B}}|\right)^2-E^2}/v\hbar\,.
\eea

\noi For every bulk segment $s=l,m,r$ the total wavefunction corresponding to energy $E$, retains the form
\bea
\Psi_{s}^E(x)=\sum_{n=1}^4c_{s;n}(E)e^{\kappa_{s;n}(E)x}\left|\kappa_{s;n}(E);\bm{{\cal B}}_s,\bm{\Delta}_s\right>\,,\qquad
\eea

\noi where the coefficients $c_{s;n}(E)$ need to be determined by imposing appropriate matching conditions at the two interfaces where the three segments meet pairwise. By ta\-king 
into account that the junction extends to infinity on the left and right parts, we conclude that the total wave-function of the system can be written as
\bea
\Psi_{l}^E(x)&=&
c_{1,l}(E)e^{+\kappa_{+}(E)(x-x_a)}\ph\left|+\kappa_{+}(E);\bm{{\cal B}}_l,\bm{\Delta}_l\right>\no\\&+&
c_{3,l}(E)e^{+\kappa_{-}(E)(x-x_a)}\ph\left|+\kappa_{-}(E);\bm{{\cal B}}_l,\bm{\Delta}_l\right>,\quad\phd\\\no\\
\Psi_{m}^E(x)&=&
c_{2,m}(E)e^{-\kappa_+(E)(x-x_a)}\left|-\kappa_{+}(E);\bm{{\cal B}}_m,\bm{\Delta}_m\right>\no\\&+&
c_{4,m}(E)e^{-\kappa_-(E)(x-x_a)}\left|-\kappa_{-}(E);\bm{{\cal B}}_m,\bm{\Delta}_m\right>\no\\&+&
c_{1,m}(E)e^{+\kappa_+(E)(x-x_b)}\left|+\kappa_{+}(E);\bm{{\cal B}}_m,\bm{\Delta}_m\right>\no\\&+&
c_{3,m}(E)e^{+\kappa_-(E)(x-x_b)}\left|+\kappa_{-}(E);\bm{{\cal B}}_m,\bm{\Delta}_m\right>,\quad\phd\\\no\\
\Psi_{r}^E(x)&=&
c_{2,r}(E)e^{-\kappa_{+}(E)(x-x_b)}\ph\left|-\kappa_{+}(E);\bm{{\cal B}}_r,\bm{\Delta}_r\right>\no\\&+&
c_{4,r}(E)e^{-\kappa_{-}(E)(x-x_b)}\ph\left|-\kappa_{-}(E);\bm{{\cal B}}_r,\bm{\Delta}_r\right>.\quad\phd
\eea

\noi The rest of the appearing coefficients in the above expressions will be determined by demanding continuity for the wave-function along the junction. There is one continuity 
equation for each interface point and they read $\Psi_l^E(x_a)=\Psi_m^E(x_a)$ and $\Psi_m^E(x_b)=\Psi_r^E(x_b)$. The arising equations define a homogeneous system of equations from 
which we may retrieve the coefficients $c_{s,n}(E)$ and additionaly the energy eigenvalues of the system.

\section{Topological - Normal - Topological (TNT) Junction}

We now proceed with examining specific junction setups. In this paragraph we shall focus on interfaces with a topological-normal-topological sequence of phases. We may recall from the 
previous section that a segment $s=l,m,r$ is in the topological phase if $|\bm{\Delta}_s|>|\bm{{\cal B}}_s|$, while the opposite condition holds for the normal phase. In order to 
obtain a double junction with the desired phase sequence per segment, we shall consider the following profile for the magnetic field and the superconducting order parameter 
$|\bm{{\cal B}}_l|=|\bm{{\cal B}}|$, $|\bm{{\cal B}}_m|=|\bm{\Delta}|$, $|\bm{{\cal B}}_r|=|\bm{{\cal B}}|$, $|\bm{\Delta}_l|=|\bm{\Delta}|$, $|\bm{\Delta}_m|=|\bm{{\cal B}}|$ and 
$|\bm{\Delta}_r|=|\bm{\Delta}|$. As a matter of fact, $|\bm{\Delta}|$ and $|\bm{{\cal B}}|$ constitute the values of the order parameter and the magnetic field in the topological 
segments. To make a connection to the results that we have obtained so far, the specific choice of values, renders only the variable 
$s_-={\rm sign}\left(|\bm{\Delta}|-|\bm{{\cal B}}|\right)$ spatially dependent, which enters in Eq.(\ref{eq:wavefunction3}) and Eq.(\ref{eq:wavefunction4}).

\begin{figure}[t]
\includegraphics[height=8.5cm,width=8.5cm]{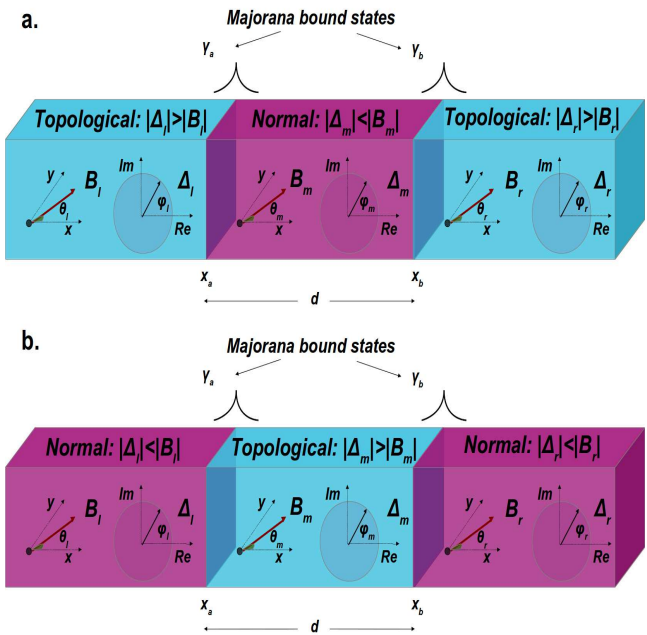}
\caption{(Color online) a. Topological-normal-topological junction, realized for the following choice of magnetic field and superconducting order parameter 
$|\bm{{\cal B}}_l|=|\bm{{\cal B}}|$, $|\bm{{\cal B}}_m|=|\bm{\Delta}|$, $|\bm{{\cal B}}_r|=|\bm{{\cal B}}|$, $|\bm{\Delta}_l|=|\bm{\Delta}|$, $|\bm{\Delta}_m|=|\bm{{\cal B}}|$ and 
$|\bm{\Delta}_r|=|\bm{\Delta}|$. b. Normal-topological-normal junction, realized for $|\bm{{\cal B}}_l|=|\bm{\Delta}|$, $|\bm{{\cal B}}_m|=|\bm{{\cal B}}|$, 
$|\bm{{\cal B}}_r|=|\bm{\Delta}|$, $|\bm{\Delta}_l|=|\bm{{\cal B}}|$, $|\bm{\Delta}_m|=|\bm{\Delta}|$ and $|\bm{\Delta}_r|=|\bm{{\cal B}}|$. In both cases localized ingap states appear at 
the interface points $x_a$ and $x_b$ where a topological quantum phase transition occurs. When the distance $d$ becomes infinite, these localized states evolve into Majorana bound 
states $\gamma_a$ and $\gamma_b$.}
\label{fig2}
\end{figure}

After applying the matching conditions for the continuity of the wave-functions  along the junction, we may obtain the bound states wave-functions and the energy eigenvalues of the 
system. Due to the particle-hole symmetry of the Bogoliubov - de Gennes Hamiltonian, we expect for every positive energy eigenvalue, an accompanying negative one. For a finite distance 
$d$ between the two junction points $x_a$ and $x_b$ (Figure \ref{fig2}), where a topological phase transition occurs, we expect the emergence of finite energy ingap bound states 
localized in the vicinity of these points. However, if the distance $d$ is taken to infinity, then the finite energy ingap states evolve into zero-energy Majorana bound states.

First we shall examine the case where there is no phase mismatch along the junction and then turn to the general case. If we set all the phases equal to zero, we find that the energy 
eigenvalues satisfy the equation $E=||\bm{\Delta}|-|\bm{{\cal B}}||e^{-\kappa_-(E)d}$. Notice that energy appears in both sides of the equation. For a large distance $d$, the energy 
tends to zero $E\rightarrow0$ and concomitantly $\kappa_-(E)\rightarrow\kappa_-(0)=||\bm{\Delta}|-|\bm{{\cal B}}||/v\hbar$. By substituting this approximate value for $\kappa_-$ back 
to the equation defining the energy, we directly obtain the expression $E=(|\bm{\Delta}|-|\bm{{\cal B}}|)e^{-(|\bm{\Delta}|-|\bm{{\cal B}}|)d/v\hbar}$, since 
$|\bm{\Delta}|>|\bm{{\cal B}}|$. We observe that the energy scales exponentially with the distance and it becomes zero for $d\rightarrow\infty$, giving rise to the zero-energy Majorana 
bound states, one per interface point. In fact, for finite $d$, the finite energy splitting is a consequence of the hybridized Majorana bound states, that now have a finite overlap. 

In order to study the case where we keep the phases of the field and the order parameter intact, we shall assume that $\kappa_{\pm}(E)\simeq\kappa_{\pm}(0)=
(|\bm{\Delta}|\pm|\bm{{\cal B}}|)/v\hbar$. To obtain an approximate analytical expression for the energy of the bound states, we shall consider that the terms $E$, $e^{-\kappa_+(0)d}$ 
and $e^{-\kappa_-(0)d}$ are of the same magnitude. In this manner, we may perform a perturbative expansion up to second order in these three terms that will provide us with the 
following compact relation for the energy
\bea
E={\cal F}(\varphi_l-\varphi_m,\vartheta_l-\vartheta_m){\cal F}(\varphi_r-\varphi_m,\vartheta_r-\vartheta_m)\qquad\no\\
\times\left[J_M\cos\left(\frac{\varphi_l-\varphi_r}{2}\right)+J_Z\cos\left(\frac{\varphi_l+\varphi_r}{2}-\varphi_m\right)\right]\,,\phd\ph
\eea 

\noi where we have defined the Josephson couplings $J_{M,Z}=\left(|\bm{\Delta}|-|\bm{{\cal B}}|\right)(e^{-\kappa_-(0)d}\pm e^{-\kappa_+(0)d})/2$ and introduced
\bea
{\cal F}(\chi,\omega)=\sqrt{\left(|\bm{\Delta}|+|\bm{{\cal B}}|\right)/\left(|\bm{\Delta}|+|\bm{{\cal B}}|\frac{\cos\chi+\cos\omega}{2}\right)}.\phd
\eea

\noi For a consistency check, we see that when the phases go to zero, ${\cal F}\rightarrow1$ and we obtain the anticipated result found earlier which yields 
$E=(|\bm{\Delta}|-|\bm{{\cal B}}|)e^{-(|\bm{\Delta}|-|\bm{{\cal B}}|)d/v\hbar}$. The term 
$J_M\cos\left(\frac{\varphi_l-\varphi_r}{2}\right)+J_Z\cos\left(\frac{\varphi_l+\varphi_r}{2}-\varphi_m\right)$ that appears in our result, is in agreement with the findings 
of Ref.\cite{Jiang}. The first coupling describes the usual $4\pi$-periodic Josephson term and the second was recently highlighted by the aforementioned authors. We observe that our 
method not only retrieves the already established results but also additional information with significant physical consequences, which are encoded in the term 
${\cal F}(\varphi_l-\varphi_m,\vartheta_l-\vartheta_m){\cal F}(\varphi_r-\varphi_m,\vartheta_r-\vartheta_m)$. Since this term includes the phases of the magnetic field 
$\vartheta_{l,m,r}$, it provides the possibility of manipulating the Josephson response via the control of the magnetic field orientation along the junction. Based on reciprocity, we 
also predict the generation of a spin current polarized along the $z-$direction, which can be controlled by tuning the superconducting phase dif\-fe\-rence along the junction. Both 
phenomena originate from the presence of the spin-orbit interaction of the semiconducting wire. 

To gain some more insight concerning these two phenomena, we shall consider the special case where $\varphi_m=(\varphi_l+\varphi_r)/2$, $\vartheta_m=(\vartheta_l+\vartheta_r)/2$ and 
$|\bm{\Delta}|>>|\bm{{\cal B}}|$. In this case, everything depends on the phase differences $\varphi_l-\varphi_r$ and $\vartheta_l-\vartheta_r$ and we obtain
\bea
E&=&\left\{1-\frac{1}{2}\frac{|\bm{{\cal B}}|}{|\bm{\Delta}|}\left[\cos\left(\frac{\varphi_l-\varphi_r}{2}\right)+\cos\left(\frac{\vartheta_l-\vartheta_r}{2}\right)\right]\right\}\no\\
&\times&\left[J_M'\cos\left(\frac{\varphi_l-\varphi_r}{2}\right)+J_Z'\right]\,,
\eea 

\noi where $J_{M,Z}'$ correspond to the values that these quantities acquire when we take the limit $|\bm{\Delta}|>>|\bm{{\cal B}}|$. We readily observe that the Josephson coupling 
consists of three types of terms: $\cos\left(\frac{\varphi_l-\varphi_r}{2}\right)$, $\cos\left(\varphi_l-\varphi_r\right)$ $\left({\rm since} \cos^2\left(\frac{\varphi_l-\varphi_r}{2}
\right)=\frac{1+\cos\left(\varphi_l-\varphi_r\right)}{2}\right)$ and $\cos\left(\frac{\varphi_l-\varphi_r}{2}\right)\cos\left(\frac{\vartheta_l-\vartheta_r}{2}\right)$. The first term 
is responsible for the $4\pi$-periodic MBS Josephson effect, the second term describes a usual $2\pi$-periodic Josephson effect and the last coupling describes a ``magnetically'' 
driven $4\pi$-periodic Josephson effect or a superconducting phase driven $4\pi$-periodic spin current. 

\section{normal - Topological - normal (NTN) Junction}

In this section we shall consider a junction consisting of a normal-topological-normal sequence of phases. To model this type of junction we shall make the following choice for the 
order parameter and the magnetic field along the junction $|\bm{{\cal B}}_l|=|\bm{\Delta}|$, $|\bm{{\cal B}}_m|=|\bm{{\cal B}}|$, $|\bm{{\cal B}}_r|=|\bm{\Delta}|$, 
$|\bm{\Delta}_l|=|\bm{{\cal B}}|$, $|\bm{\Delta}_m|=|\bm{\Delta}|$ and $|\bm{\Delta}_r|=|\bm{{\cal B}}|$. Once again, $|\bm{\Delta}|$ and $|\bm{{\cal B}}|$ correspond to the values of 
the superconducting gap and Zeeman field in the topological region. The methodology we use is identical to the one followed in the previous section and we obtain the expression for the 
midgap state energy 
\bea
E={\cal F}(\varphi_l-\varphi_m,\vartheta_l-\vartheta_m){\cal F}(\varphi_r-\varphi_m,\vartheta_r-\vartheta_m)\qquad\no\\
\times\left[J_M\cos\left(\frac{\vartheta_l-\vartheta_r}{2}\right)+J_Z\cos\left(\frac{\vartheta_l+\vartheta_r}{2}-\vartheta_m\right)\right]\,,\phd\ph
\eea 

\noi where we have used the same definitions as previously. We observe that the results of this section can be related to the prior ones by just exchanging the phases 
$\varphi\leftrightarrow\vartheta$. Again, if we take the special case $\varphi_m=(\varphi_l+\varphi_r)/2$, $\vartheta_m=(\vartheta_l+\vartheta_r)/2$ and the limit
$|\bm{\Delta}|>>|\bm{{\cal B}}|$, we similarly obtain
\bea
E&=&\left\{1-\frac{1}{2}\frac{|\bm{{\cal B}}|}{|\bm{\Delta}|}\left[\cos\left(\frac{\varphi_l-\varphi_r}{2}\right)+\cos\left(\frac{\vartheta_l-\vartheta_r}{2}\right)\right]\right\}
\no\\
&\times&\left[J_M'\cos\left(\frac{\vartheta_l-\vartheta_r}{2}\right)+J_Z'\right]\,.
\eea 

\noi In this case we also retrieve three different kinds of couplings $\cos\left(\frac{\vartheta_l-\vartheta_r}{2}\right)$, $\cos\left(\vartheta_l-\vartheta_r\right)$ and 
$\cos\left(\frac{\varphi_l-\varphi_r}{2}\right)\cos\left(\frac{\vartheta_l-\vartheta_r}{2}\right)$. The first two terms are connected to $4\pi$-periodic and $2\pi$-periodic spin 
transport, while the novel type of ``magnetic'' Josephson coupling (third term) still appears in this case.

\section{Magnetically tuned Josephson coupling and alternative way of Braiding}

Apart from the importance of exploring routes for engineering MBS and topological qubits, an equally significant task is to invent new techiques for manipulating them and performing 
quantum computations. MBS offer due to their topological stability significant advantages as compared to the more conventional qubits based on either Josephson junctions 
\cite{Schoen} or spins in quantum dots \cite{DiVincenzo}. However, for their manipulation new techniques have to be developed. One of the standard methods in the latter situation, is to create 
junctions (e.g. Y-junctions \cite{Sau Majo exchange,Halperin}) where a number of MBS are brought together, interact and afterwards are separated again in order for a read-out 
protocol to be implemented and yield the quantum computation output. Up to now, the only interaction considered to take place between two MBS was through the Josephson coupling 
\bea
{\cal H}_{M}\propto i\gamma_l\gamma_r\cos\left(\frac{\varphi_l-\varphi_r}{2}\right)\,.
\eea 

\noi As a matter of fact, the superconducting phase of the order parameter has to be manipulated in order to braid the MBS. However, in this paper we revealed another type of 
Josephson coupling that exists in these systems and this is of the form
\bea
{\cal H}_{M}^{mag}\propto i\gamma_l\gamma_r\cos\left(\frac{\vartheta_l-\vartheta_r}{2}\right)\cos\left(\frac{\varphi_l-\varphi_r}{2}\right)\,.
\eea

\noi This novel Josephson coupling provides an additional degree of freedom to be manipulated, which is the orientation of the Zeeman magnetic field. In this manner, varying the 
superconducting order parameter phases is not necessary any more, since braiding can be performed by properly adjusting the magnetic field that each MBS feels.

\section{CONCLUSIONS}

We presented a detailed analysis of the Josephson effect in double junction sandwiches of topologically trivial and non-trivial phases of semiconductor-superconductor 1D 
heterostructures. We demonstrated the existence of a novel type of Josepshon coupling that allows the manipulation of the supercurrent via spatial variations of the Zeeman field
along the junction. Conversely, a spin current generation is possible by varying the superconducting order parameter phase differences. This additional term opens a pathway for 
unambiguous Majorana bound state detection and at the same time constitutes an alternative platform for topological qubit operations. 

\bt{\textit{Note added:}} After we had already submitted this manuscript for the ICM2012 proceedings, two other papers \cite{Meng,Pekker} appeared, arriving independently to similar 
conclusions.

\begin{acknowledgments}
We are grateful to J. Michelsen, P.-Q. Jin, S. Andr\'{e}, R. Grein and A. Heimes for comments and stimulating discussions. We further acknowledge financial support from the EU 
projects NanoCTM, SOLID and GEOMDISS.
\end{acknowledgments}

\end{document}